\documentclass{optica-article}

\journal{opticajournal} 

\articletype{Research Article}

\usepackage{lineno}

\hypersetup{
 colorlinks=true,
 linkcolor=blue,
 citecolor=blue}
 
\begin{document}

\title{FPM-WSI: Fourier ptychographic whole slide imaging via feature-domain backdiffraction}

\author{Shuhe Zhang,\authormark{1,2,3,$\dag$} Aiye Wang,\authormark{1,4,$\dag$} Jinghao Xu,\authormark{1,4} Tianci Feng,\authormark{1,4} Jinhua Zhou,\authormark{3} and An Pan\authormark{1,4,*}}

\address{\authormark{1}State Key Laboratory of Transient Optics and Photonics, Xi’an Institute of Optics and Precision Mechanics, Chinese Academy of Sciences, Xi’an 710119, China\\
\authormark{2}Maastricht University Medical Center +, Maastricht, 6202 AZ, the Netherlands\\
\authormark{3}School of Biomedical Engineering, Anhui Medical University, Hefei 230032, China\\
\authormark{4}University of Chinese Academy of Sciences, Beijing 100049, China\\
\authormark{\dag}The authors contributed equally to this work.}
\email{\authormark{*}panan@opt.cn} 


\begin{abstract*} 
Fourier ptychographic microscopy (FPM), characterized by high-throughput computational
imaging, theoretically provides a cunning solution to the trade-off between spatial resolution and field of view (FOV), which has a promising prospect in the application of digital pathology. However, block reconstruction and then stitching has currently become an unavoidable procedure due to vignetting effects. The stitched image tends to present color inconsistency in different image segments, or even stitching artifacts. Consequently, the advantages of FPM are not as pronounced when compared to the conventional scanning-and-stitching schemes widely employed in whole slide imaging (WSI) systems. This obstacle significantly impedes the profound advancement and practical implementation of FPM, explaining why, despite a decade of development, FPM has not gained widespread recognition in the field of biomedicine. In response, we reported a computational framework based on feature-domain backdiffraction to realize full-FOV, stitching-free FPM reconstruction. Different from conventional algorithms that establish the loss function in the image domain, our method formulates it in the feature domain, where effective information of images is extracted by a feature extractor to bypass the vignetting effect. The feature-domain error between predicted images based on estimation of model parameters and practically captured images is then digitally diffracted back through the optical system for complex amplitude reconstruction and aberration compensation. Through massive simulations and experiments, the method presents effective elimination of vignetting artifacts, and reduces the requirement of precise knowledge of illumination positions. We also found its great potential to recover the data with a lower overlapping rate of spectrum and to realize automatic blind-digital refocusing without a prior defocus distance. Furthermore, to the best of our
knowledge, we firstly demonstrated application of FPM on a WSI system, termed FPM-WSI. This platform
enables full-color, high-throughput imaging (4.7 mm diameter FOV, 336 nm half-pitch resolution with
blue channel illumination) without blocking-and-stitching procedures for a batch of 4 slides. The platform also possesses autofocusing, shifting and regional recognition of slides that are completed by additional automatic mechanical hardware, and the acquisition time for a single slide is less than 4 s. In addition, we provide a user-friendly operation interface to facilitate the workflow, and alternative colorization schemes to choose from. The impact of the reported platform, with advantages of high-quality, high-speed imaging and low cost, will be far-reaching and desired in many fields of biomedical research, as well as in clinical applications.

\end{abstract*}

\section{Introduction}
For decades, using the conventional optical microscopy for pathological analysis has been the gold standard of disease detection and grading, where pathologists assess several tissue slides to attain precise observation of cellular features and growth pattern. Ruling out subjective factors, the accuracy of diagnosis profoundly depends on the throughput of imaging system. Hence, high-throughput microscopic imaging is of great significance for the study of pathological mechanisms and effective therapy of diseases, which is also intensively explored in applications like haematology \cite{zhu2022real}, immunohistochemistry and neuroanatomy \cite{lu2022somatosensory,felger2023robustness,banerjee2023analogous}.  

The throughput of an optical imaging system is determined fundamentally by its space-bandwidth product (SBP), which is defined as the number of resolvable pixels in the imaging field of view (FOV). However, the achievable SBP is in essence restricted by the scale-dependent geometric aberrations of the optical elements, leading to a trade-off between image resolution and FOV. A natural solution, from the point of optical design, is to optimize the aberrations caused by large-scale elements, but the resultant utilization of multiple lens considerably escalates the system volume and complexity. The demand for high SBP microscopic systems in the field of pathology and biomedicine has spurred the development and commercialization of whole slide imaging (WSI). Instead of manually examining glass slides with a microscope eyepiece, WSI digitalizes the entire FOV of a histological or biological specimen at high resolution (HR) for pathologists, researchers and clinicians to observe and analyze on a computer screen \cite{farahani2015whole,kanwal2022devil}. The workflow of existing WSI systems generally consists of two parts: the first entails a specialized high-precision scanner to capture a series of HR images corresponding to different regions of the slide; the second stitches together these image segments into a full-FOV image of the slide by professional softwares. However, inevitable errors in mechanical scanning easily cause misalignment in the stitched image despite a sufficient overlapping rate. Uneven illumination of light source will also lead to uneven distribution of brightness or even stitching stripe-artifacts, which not only deteriorates the quality of stitched image, but also affect the quantitative analysis of downstream applications. For example, one study suggested that ignoring illumination correction resulted in a 35\% increase of false and missed detections of yeast cells images \cite{smith2015cidre}. Both conventional \cite{smith2015cidre} and deep-learning \cite{wang2023deep} methods attempt to eliminate the artifacts based on post processing, not yet tackling the problem fundamentally.

Inspired by the concept of synthetic aperture \cite{moreira2013tutorial,holloway2017savi}, Fourier ptychographic microscopy (FPM) provided a brand-new perspective for the seeking of high-SBP microscopic systems \cite{zheng2013wide}. With the multiple-angle illumination of an LED array, FPM acquires corresponding low resolution (LR) images and stitches them in the Fourier domain to reconstruct an HR complex amplitude image of the sample. As the used low numerical aperture (NA) objective has an innate large FOV, FPM enables high-SBP imaging without mechanical scanning, and thus can bypass the artifacts caused by image stitching. In practical implementations, however, full-FOV FPM reconstruction highly relies on block processing \cite{pan2019vignetting,gao2023design}. The vital consideration for doing so is to avoid the degraded reconstruction quality caused by vignetting effect, typically severe wrinkle-artifacts appearing at the edge of FOV. The requirement of plane wave illumination \cite{zheng2016fourier}, coherence and reducing computational load also makes block processing almost a compulsory option. However, the resultant reconstruction still suffers from digital stitching artifacts (distinguished from mechanical ones). That’s because conventional reconstruction algorithms require high precision of parameters, and inadequate noise removal \cite{yeh2015experimental,zuo2016adaptive,claveau2020structure} or system error-correction (e.g. deviation of illumination positions \cite{eckert2018efficient,zhang2019positional,huang2020positional}, intensity fluctuations of LEDs \cite{bian2013adaptive,hou2018adaptive}) will cause color inconsistency of different reconstructed image segments. Some methods try to avoid block reconstruction by solving vignetting at the hardware level, such as designing an illumination source with special LED layout \cite{zhu2020single,gao2023design}, which undoubtedly increases the system complexity. Solving the requirement of parameters at the algorithmic level might be feasible. Unfortunately, the models of existing methods are all established based on image-domain optimization \cite{yeh2015experimental,zheng2021concept}, in which the loss function for FPM reconstruction is designed to calculate the difference between estimated images and experimentally captured images. They have to face the challenges of noise signals and systematic errors, because these factors of image degradation cannot be split away from the image domain. The majority of studies, therefore, focused on the display of subregion reconstruction. Some exceptions \cite{zheng2013wide,tian2014multiplexed,valentino2023beyond} have reported their impressive performance of stitched full-FOV reconstructions, while the mentioned flaw can still be slightly detected. This fundamentally explains why the advancement and landing of FPM in digital pathology faces numerous obstacles, and why FPM has not been widely accepted by the field of biomedicine even with a 10-year development.

In this paper, we reported a direct, non-blocked full-FOV reconstruction method of FPM based on feature-domain back diffraction. Unlike previous algorithms, the loss function for reconstruction is uniquely formulated in the feature domain of images. As the vignetting effect has a slow variation characteristic in the feature domain, the effective information can be extracted from the degraded images with a certain designed feature extractor \cite{kimmel2003variational,li2010selection,zhang2023elfpie}, and the undesired influence of vignetting effect can be bypassed.The feature-domain error is then digitally diffracted back through the optical system for complex amplitude reconstruction and aberration compensation. Intensive simulations and experiments have verified that our method can fundamentally remove vignetting artifacts and thus omit blocking-and-stitching procedures. The precision requirement of LED positions for FPM systems becomes relaxed, even raw data acquired using an even-number illumination array can be well reconstructed. 

Interestingly, our method also presents impressive performance in recovering data with a lower overlapping rate of spectrum and completing  digital refocusing without prior knowledge of defocus distance. Furthermore, we firstly reported in this paper on applying FPM to a WSI system, named FPM-WSI. On the hardware side, the proposed WSI platform is adapted from a conventional trinocular inverted microscope. The high-brightness LED source enables only 2ms exposure for each capture of the camera. A $z$ axis driver and an $x$-$y$ axis electric displacement stage are respectively equipped for autofocusing and for automatic shift of batch samples (4 slides). On the software side, the platform provides optional colorization schemes, precise autofocusing method and user-friendly operation interface. The reported system is expected to break the bottleneck that has long been constraining the development of FPM, which will offer a ingenious solution to transform WSI platforms into ones that can be broadly accepted and utilized in the field of biomedical research and clinical applications.

\section{Results}
\begin{figure}[ht!]
\centering\includegraphics[width=0.95\textwidth,trim = 0 100 0 0,clip]{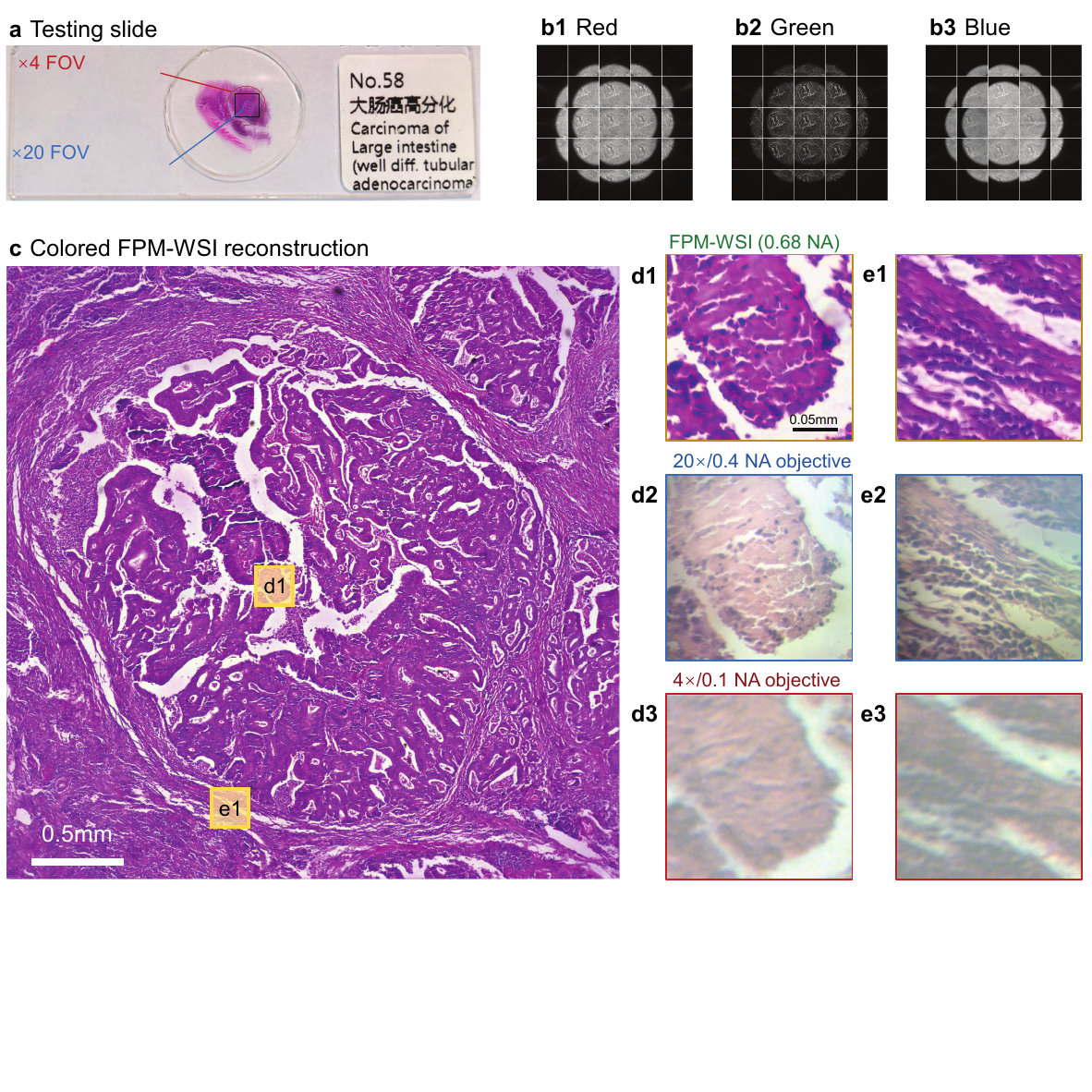}
\caption{Color image of a pathology slide via FPM-WSI. (a) sample slide. (b1), (b2) and (b3) show first 25 images of FPM datacube for red, green and blue colors illumination. Vignetting effect is prominent. (c) $3.3 \times 3.3 \text{ mm}^2$ Full-FOV image of a human colorectal carcinoma section created by fusing reconstructions of R/G/B channels (refer to \url{https://www.gigapan.com/gigapans/233966}); (d1) and (e1) are reconstructions of two ROIs marked in (c), with a size of $650 \times 650$ pixels; (d2,e2) and (d3,e3) are corresponding images captured by a color image sensor using a $\times 20$ and $\times 4$ objective lens for comparison. \label{fig1_results}}
\end{figure}

Fig. \ref{fig1_results} (c) demonstrates the wide-FOV color image of a pathology slide [human colorectal carcinoma section in Fig. \ref{fig1_results} (a)] by directly fusing FPM reconstructed results of three color channels despite the presence of prominent vignetting effect, as shown in Fig. \ref{fig1_results} (b). Fig. \ref{fig1_results} (d1,e1) present a magnified view of two regions of interest (ROI). Shifting the light source to a halogen lamp, we also utilized a color charge coupled device (CCD) camera (ImagingSource DFK 23U445) to capture the images of corresponding regions with a 20×/0.4NA and a 4×/0.1NA objective lens for comparison, as shown in Fig. \ref{fig1_results} (d2,e2) and Fig. \ref{fig1_results} (d3,e3). The maximum synthetic NA for this experimental setup achieves 0.68 set by the angle between the optical axis and the LED located at the outermost edge. Based on the concept of synthetic aperture, FPM realizes a resolution improvement of approximately seven times compared with simply imaging via a 0.1NA objective. Theoretically, the resolution of FPM reconstruction is close to that of 0.4NA incoherent imaging (equivalent synthetic NA of 0.8). As to the contrast of Fig. \ref{fig1_results} (e1,e2), even when the observed region has a dense distribution of cells, the details can be clearly identified in both results. However, the FPM reconstruction provides better contrast of display than the result of 0.4NA imaging.

The most attractive elegance of FPM lies in large-FOV imaging without mechanical scanning, while the proposed feature-domain-based reconstruction method further removes the necessity of block processing used in conventional gradient descent iterative algorithms. The image is free of stitching artifacts and color differences as no segmentation of each full-FOV raw image was applied. We also do not find typical vignetting artifacts at the image margin, verifying the unique effect of our method on artifact elimination. For comparison, the result of scanning and stitching and reconstructions via gradient descent method (AS-EPRY: EPRY \cite{ou2014embedded} with adaptive step-size \cite{zuo2016adaptive}) for the same slide are respectively discussed in \textbf{Supplementary 1} and \textbf{Supplementary 2}.

\section{Material and methods}

\subsection{FPM-WSI platform}

\begin{figure}[ht!]
\centering\includegraphics[width=0.82\textwidth,trim = 0 0 250 0,clip]{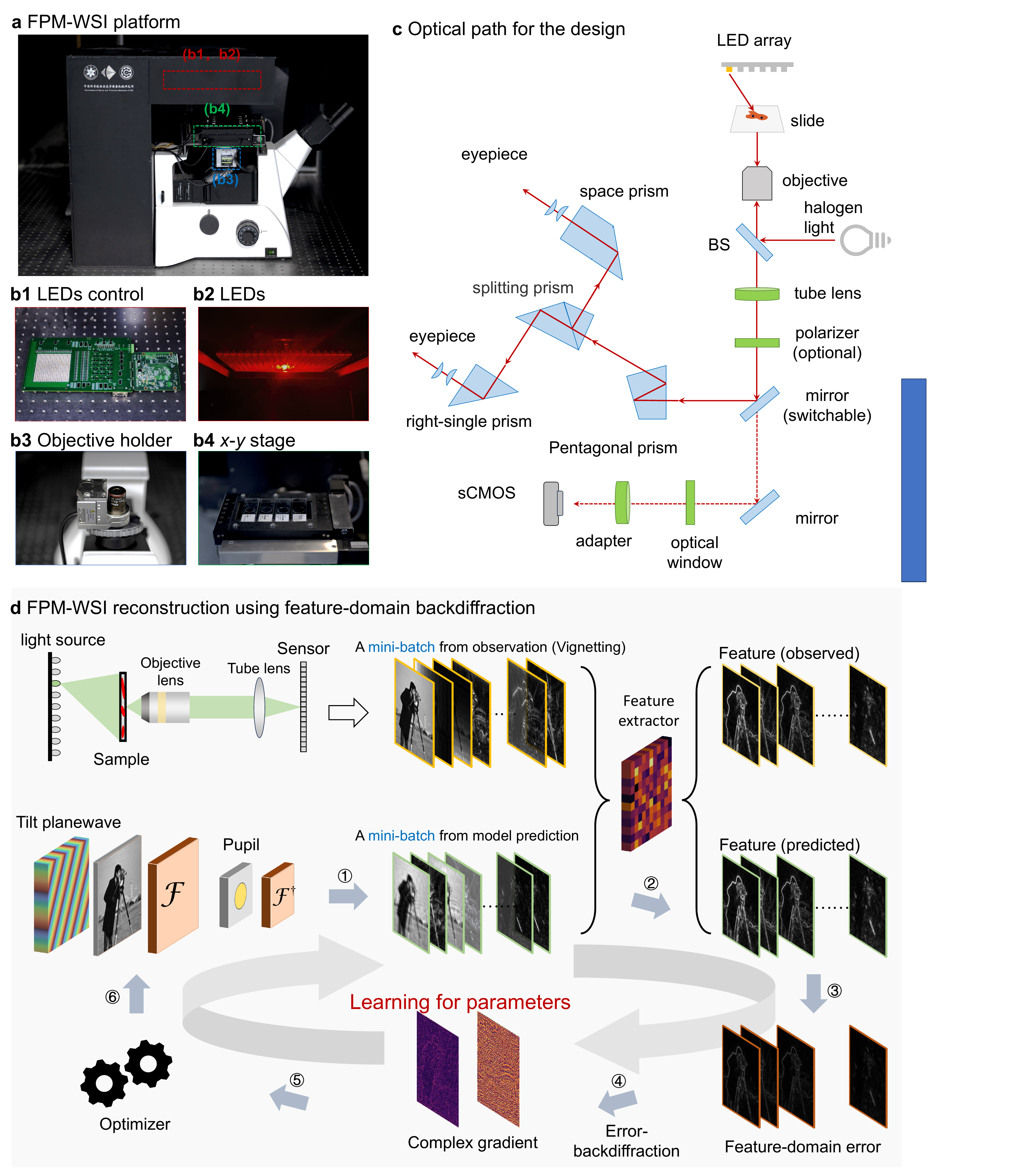}
\caption{FPM-SIM platform setup. (a) Overall architecture of FPM-SIM platform generally consisting of microscopic imaging system, automatic control system and a host computer; (b1) $19 \times 19$ programmable LED array for sample illumination; (b2) Packaged appearance of LED array with the central LED lightning; (b3) $z$ axis driver holding the objective lens for autofocusing; (b4) $x-y$ axis displacement stage for mechanical movement of a batch of 4 slides; (c) Optical path diagram of the microscope. (d) Flowchart for FPM-WSI reconstruction using feature-domain backdiffraction, involving 6 steps. Step 1: the model generates a series of predicted images based on current estimation of parameters including the comlex amplitude of the sample and pupil function. Step 2: the predicted images and their corresponding observed images are filtered by the feature extractor, producing the feature maps. Step 3: the feature-domain error between model prediction and the observations is calculated. Step 4: the error is back-diffracted to yield the complex-gradient. Step 5: the complex-gradient is managed by the optimizer with potential first-order and second-order moments. Step 6: the model parameters are updated. \label{fig2_method}}
\end{figure}

Fig. \ref{fig2_method} (a) shows the system integration of our high-throughput automatic WSI platform, and Fig. \ref{fig2_method} (b1-b4) present the details of different components marked in Fig. \ref{fig2_method} (a). \textbf{Visualization 1} gives the overall display of the platform. The platform can generally be divided into four parts: illumination source, automatic control system, main body of the microscopic imaging system and a host computer.

The illumination source for FPM should meet the basic requirements of high brightness and high refreshing rate, in order to reduce the time for data acquisition. Accordingly, we designed a programmable LED array containing 19×19 surface-mounted full-color LEDs [Fig. \ref{fig2_method} (b1)], and the distance between two adjacent LEDs is 4 mm (referred to Section 4.4 for the selection of LED parameters). The central wavelengths of three color channels are 631.23 nm (red), 538.86 nm (green) and 456.70 nm (blue), each offering an approximately spatially coherent quasi-monochromatic source with 20 nm bandwidth. We used LM3549 driver chips to provide the logical control for the LED array. For each LED, the measured maximum power of a single channel is 1 W, and the refreshing rate is no less than 100 Hz. In practical use with a 16-bit sCMOS camera, the exposure time for both brightfield and darkfield images can be controlled properly at 2 ms. The frame rate of our camera is 100 fps, resulting in an acquisition time of 10 ms for each raw image and totally less than 4 s for a single slide (see \textbf{Visualization 2}). There is still a large space for the reduction of acquisition time if employing a camera with higher frame rate. Notably, as shown in Fig. \ref{fig2_method} (b2), the LED array is packaged into an opaque sealed environment, except that the side facing the slide is exposed to the surroundings. Together with the high-brightness illumination, the majority of stray light can be suppressed, and there is no need for the platform to work under strict darkroom conditions. Compared with our previous work on monochromatic hemispherical illuminator \cite{pan2018subwavelength}, besides the significant improvement of acquisition efficiency, the manufacture difficulty of this R/G/B source is greatly increased. Even so, the flat-panel structure also has tremendous advantages in standardization and pipeline production.

The automatic control system consists of a $z$ axis driver [Fig. \ref{fig2_method} (b3)] and an $x$-$y$ axis electric displacement stage [Fig. \ref{fig2_method} (b4)]. The operation of the system can be seen in \textbf{Visualization 3}. The $z$ axis drive (OptoSigma, SGSP-OBL-3) is used to control the mechanical movement of the objective lens for autofocusing. The drive covers a range of 3 mm with the resolution of 1 $\upmu$m/pulse, basically satisfying the focusing demand of slides with various thicknesses. The command transmission between the drive and the host computer is realized via a GIP-101B controller. The $x$-$y$ axis electric displacement stage enables precise positioning and automatic shift between a batch of 4 slides (25 mm × 75 mm), thus we customized a rectangle aluminum alloy plate embedded with 4 slide slots and fixed it on the upper surface of the displacement stage. The range of the stage at two directions is 120 mm and 50 mm respectively, both with 10 $\upmu$m repetition precision. 

The main body of the microscopic imaging system was adapted from a conventional trinocular inverted microscope, whose optical path is demonstrated in Fig. \ref{fig2_method} (c). The trinocular design supports wide-angle observation through the eyepiece. In addition, the system is highly flexible with expansibility. A built-in halogen light source allows us to switch between the FPM imaging mode and the regular bright-field imaging mode. Other microscopy techniques such as polarization imaging and fluorescence imaging can also be implemented using this system as the polarizer and ultraviolet lamp are optional to be equipped. 

The host computer with an Intel i5 CPU, 32 GB RAM is mainly responsible for data storage and processing, control of automatic devices and user access. This configuration averts the high cost associated with modifying, manufacturing and maintaining a GPU-based device, which appeals more to healthcare organizations and research institutes. Here, we would like to highlight three characteristics in terms of software. First, different from conventional WSI platforms, our system does not adopt focus map-based methods \cite{montalto2011autofocus} to realize autofocusing. Instead, we propose a hill-climbing algorithm where two symmetric LED units with different wavelengths light up and the degree of focus is determined based on evaluation of spectrum energy (see \textbf{Supplementary Note 3}). Second, as we all know, color FPM images can be created by simply combining reconstructed results from R/G/B LED illumination into each corresponding color channel. Our WSI platform additionally incorporates a state-of-art colorization method named color-transfer filtering FPM (CFFPM) \cite{chen2022rapid} as an alternative option, which sacrifices minimal precision imperceptible to human vision while tripling the acquisition efficiency. Third, all software modules of the platform have been integrated into Matlab 2022, including data acquisition, LED control, autofocusing, slide shifting, FPM reconstruction and colorization. For the refinement and controllability of experiments, we also designed a user-friendly operation interface to facilitate the implementation of a complete workflow. (see \textbf{Supplementary 4}) 

\subsection{Feature-domain backdiffraction}

The forward model of FPM reads:
\begin{equation}
\mathbf{I}_n = \left | \mathbf{F^{\dagger}PM}_n\mathbf{FU} \right |^2 + \epsilon, \label{Eq. 1}
\end{equation} 
where $\mathbf{I}_n$ is the $n$-th captured low-resolution (LR) images illuminated by corresponding LED. $\mathbf{U}$ is the complex amplitude of the object, and $\mathbf{F}$ denotes the Fourier transform (Fraunhofer diffraction) when $\mathbf{U}$ propagates through the objective lens. The superscript symbol $\dagger$ means the conjugate transpose of a matrix, and $\mathbf{F} ^\dagger$ denotes the inverse Fourier transform since $\mathbf{F} ^\dagger = \mathbf{F} ^{-1}$. $\mathbf{M}_n$ is the selection matrix for the $n$-th LED of the array. $\mathbf{P}$ is the pupil function of the imaging system.  

FPM reconstruction can be regarded as a maximum a posteriori estimate (MAP) problem \cite{gribonval2011should,pereyra2017maximum} in which we are going to find an estimation of parameter that can best explain the observed data through the forward model. Conventional reconstruction methods are based on the ptychographic iterative engine (PIE) that maximizes the Gaussian-likelihood, or in other words, minimizes the $L_2$-distance (Euclidean distance) between model prediction and observations in the image domain \cite{yeh2015experimental}, given as $\mathcal{L}_{\text{Conventional}}\left ( \mathbf{U},\mathbf{P} \right )  =  \left \| \sqrt{\mathbf{I}_n}- \left | \mathbf{F^{\dagger}PM}_n\mathbf{FU} \right |  \right \| ^2_2.$ The reconstruction results are highly susceptible to vignetting effect, noise signals and systematic errors. Some studies tried to maximize the Poisson-likelihood \cite{bian2016fourier}, yet still cannot address the vignetting effect as the root of problem lies in the mismatch of forward model. In fact, Eq. (\ref{Eq. 1}) holds for an approximated linear space-invariant (LSI) coherent microscopic system, where the coherent transfer function (CTF) is determined by a complex pupil function of the objective. However, it is not well suited to model the half-bright and half-dark vignetting effect in practical experimental conditions, as a typical FPM imaging system cannot be classified as a strict LSI system without the introduction of complicated Fresnel diffraction theory.

Instead of adopting a more complicated forward model, our FPM-WSI minimizes the $L_1$-distance (Manhattan distance) on the feature domain of images [Fig. \ref{fig2_method} (d)], and the feature-domain loss function is given as
\begin{equation}
\mathcal{L}_{\text{FPM-WSI}}\left ( \mathbf{U},\mathbf{P} \right )  = \sum_{n=1}^{N} \left \| \mathcal{K} \sqrt{\mathbf{I}_n}-\mathcal{K} \left | \mathbf{F^{\dagger}PM}_n\mathbf{FU} \right |  \right \| _1, \label{Eq. 2}
\end{equation}
where $\mathcal{K}$ denotes the invertible convolution kernel for feature extraction. In this implementation, we use the first-order image's edge to represent the feature and $\mathcal{K} = \left[ {\nabla}_x,{\nabla}_y \right]^\top$. The $L_1$-distance is derived from the statistical fact that the edge of image follows heavy-tail distribution \cite{kotera2013blind}, which can be approximated by Laplacian distribution. Moreover, $L_1$-distance promotes the sparsity of input vector which favors the edge features \cite{candes2008enhancing}. Although $L_1$-norm is non-differentiable at the origin, the cost function in this case is zero and there is no need to update the parameters. Therefore, we are able to use the sub-gradient for parameter learning. The benefits of proposed feature-domain loss function is detailed explained in Subsec. \ref{subsec4.1}.

The gradient of Eq. (\ref{Eq. 2}) w.r.t. $\mathbf{U}$ and $\mathbf{P}$ are calculated by $\mathbb{CR}$-calculus \cite{kreutz2009complex}, and then  digitally diffracted back through the optical system for parameter update,  including the complex amplitude of the object and the CTF. The details of this process are given in \textbf{Supplementary 5}. The back-diffraction in optics literally refers to propagating the wavefront to the object plane as sketched in Fig. \ref{fig2_method} (d). For example, the Fraunhofer propagator that diffracts the input wave to the output wave is described as the Fourier transform, while its Hermitian transpose denotes the inverse Fourier transform which exactly diffracts the output wave back to the input wave. 

It is worth to mention that $N$ in Eq. (\ref{Eq. 2}) can be any integer ranging from 1 to the total number of raw images. This flexibility allows for the computing batch gradients from randomly selected images from the raw image dataset. While employing all raw images for gradient calculation leads to a global gradient descent, utilizing a smaller batch of raw images can enhance the possibility of the algorithm escaping local minima, especially given the severely non-convex nature of the optimization process.

\section{Discussions}
\subsection{Feature-domain loss and batch gradient descent} \label{subsec4.1}

The feature-domain loss, based on the first-order edge of image, effectively suppresses the impact of vignetting effect due to the different statistical properties between image domain and feature domain. Notably, the vignetting effect, which lacks sharp edges, manifests solely in the image domain but is absent in the feature domain, meaning that we can efficiently separate them in image's edge-feature domain. Let $\sqrt{I_\text{ideal}}$ and $\sqrt{I_\text{vignet}}$ be the amplitude in an ideal condition and in the case of vignetting. Simply minimizing the $L_2$-distance between them is ineffective and the discrepancy remains substantial, as the forward model in Eq. (\ref{Eq. 1}) fails to encapsulate the vignetting effect. In other words, the MAP cannot find a good estimation of model parameter that can explain the observations. This situation introduces unexpected low-frequency components into the reconstructed Fourier spectrum, which is reflected as severe wrinkle artifacts.

\begin{figure}[ht!]
\centering\includegraphics[width=0.85\textwidth,trim = 0 50 0 0,clip]{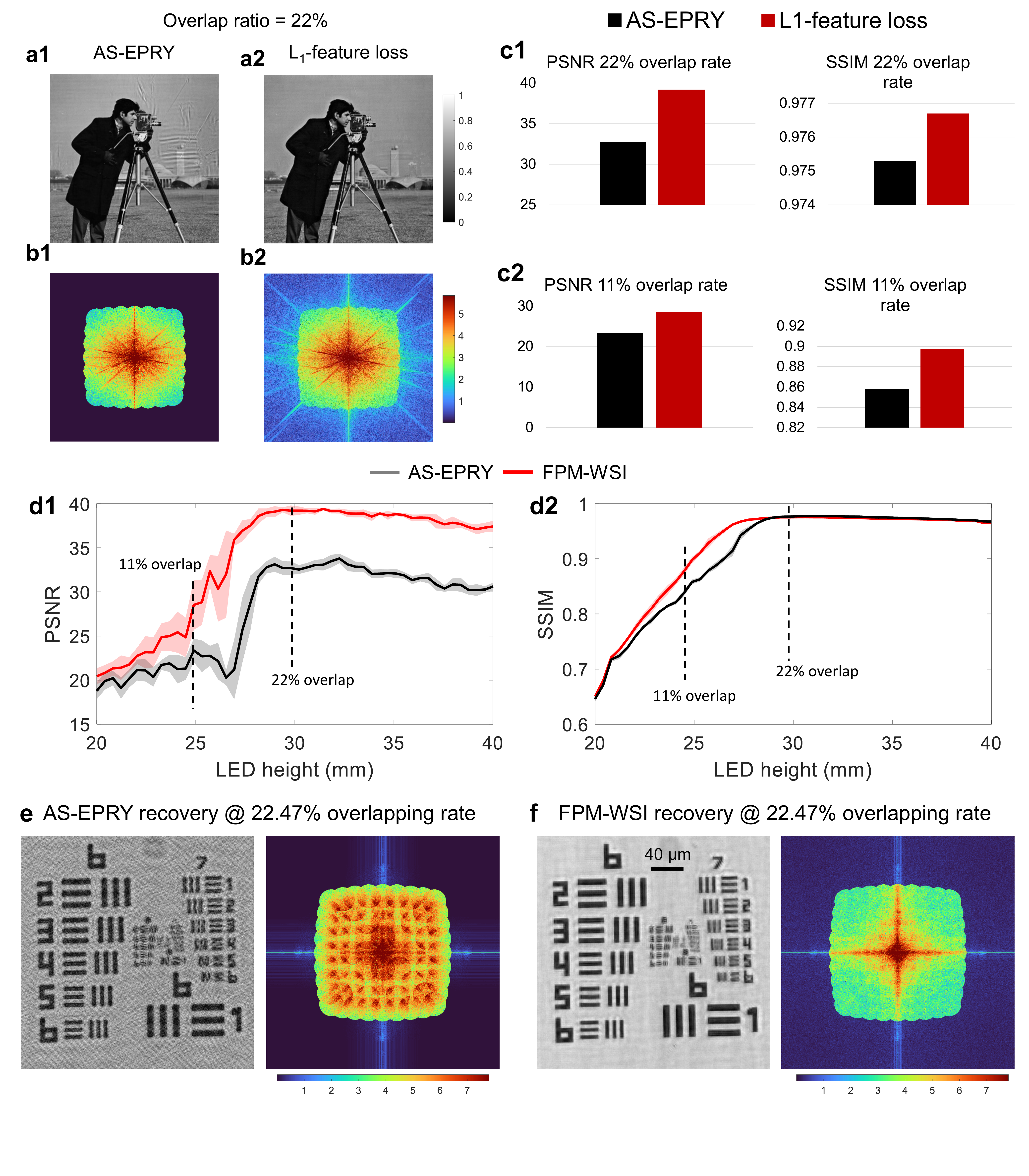}
\caption{Reconstructions for lower overlapping rate of spectrum. (a1-a2) and (b1-b2) are reconstructed amplitudes and their spectrum with simulated overlapping rate of 22\%. (c1-c2) lists the average value of PSNR and SSIM for two simulations of overlapping rates. (d1) and (d2) plot PSNR and SSIM score for 500 groups of simulation study, 2\% Gaussian noise was added. The LED height controls the overlapping rates. (e) and (f) show experimental results using AS-EPRY and FPM-WSI respectively when the overlapping rate of spectrum is 22.47\%. \label{fig3_overlap}}
\end{figure}

While taking the first-order spatial gradient of predicted and observed data, the distance between $\nabla \sqrt{I_\text{ideal}}$ and $\nabla \sqrt{I_\text{vignet}}$ becomes very close because the vignetting effect has a slow spatial variation that contributes gentle spatial gradient values compared to that of valid sample structures. More importantly, the edge information of an image is sparsely distributed and can be approximate to Laplacian distribution \cite{levin2011understanding}. Consequently, minimizing the $L_1$-distance in Eq.(\ref{Eq. 2}) provides an effective solution to circumventing the troublesome vignetting effect. Moreover, the $L_1$-distance takes more efficient use of data information, thus enabling robust FPM reconstruction even at a low overlapping rate of spectrum. According to the redundant information model for FPM \cite{gao2023redundant}, the utilization rate of our method is approximately calibrated as 30\%, higher than that of conventional gradient descent algorithms (24\%). Previous work \cite{sun2016sampling} indicated that at least 35\% overlapping rate of sub-apertures in the Fourier domain is required for a successful reconstruction using conventional FPM algorithms. However, we demonstrated that our method could break through the lower limit. We created two groups of simulated data with overlapping rate of spectrum down to 22\% and 11\%. The AS-EPRY fail to reconstruct the amplitude in high quality, generating obvious crosstalk with phase information, as shown in Fig. \ref{fig3_overlap} (a1-b2). On the contrary, FPM-WSI with feature-domain loss still works promisingly as indicated by the comparison of structure similarity (SSIM) and peak signal-to noise ratio (PSNR) in Fig. \ref{fig3_overlap} (c,d). We also experimentally examined the reconstruction performance of AS-EPRY and FPM-WSI, and the result of comparison is shown Fig. \ref{fig3_overlap} (e,f). Here, we set the illumination height as 30 mm, and the overlapping rate of spectrum is calculated as 22.47\%. Merely from a visual perspective, the same conclusion can be drawn as in the simulations.

Different from conventional techniques, which frequently update the Fourier spectrum of object for each individual illumination angle, FPM-WSI incorporates batch-based gradient descent that computes the collective gradient for a mini-batch involving multiple illumination angles. This approach allows compensation of gradient components in the overlapping area of sub-apertures, resulting in acceleration of convergence and also showing robustness to data noises. Before utilizing the gradient to update spectrum, the processing of optimizer such as Adam \cite{kingma2014adam}, RMSprop or YOGI \cite{zaheer2018adaptive} is necessary. The optimizer generates the first-order and second-order moment, which can not only accelerate the convergence of this non-convex phase retrieval problem, but also facilitate the update of spectrum exceeding the synthetic aperture, as illustrated in Fig. \ref{fig3_overlap} (b2,b3) and Fig. \ref{fig3_overlap} (d2,d3).

\subsection{Experimental robustness}

\begin{figure}
\centering
\includegraphics[width = 0.90\textwidth,trim = 0 350 0 0,clip]{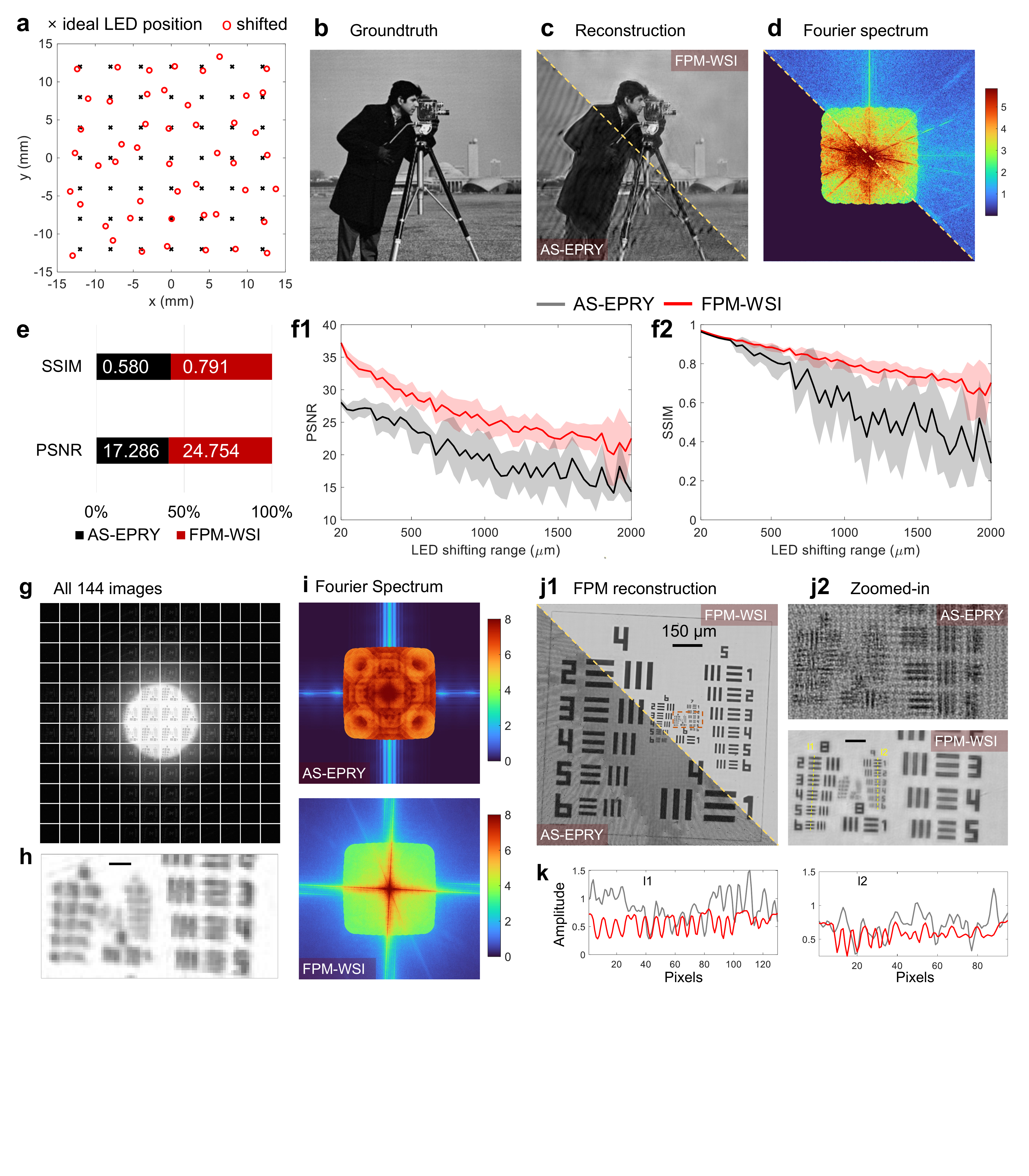}
\caption{Comparison of experimental robustness for conventional FPM algorithm and FPM-WSI. (a) Simulated LED position shift in the LED array. (b-d) are simulated ground truth, reconstructed amplitude and corresponding Fourier spectrum. (e) lists the value of PSNR and SSIM for two methods. (f1,f2) plot the value of PSNR and SSIM for 500 groups of simulation with different degrees of LED position shift. (g) Raw data obtained with the illumination of 12 × 12 LEDs, and obvious vignetting effect can be found in the central 4 × 4 images. (h) Magnified view of ROI in the raw data. (i) Fourier spectrum of reconstruction using AS-EPRY and FPM-WSI. (j1,j2) show the reconstructed amplitude of USAF target and magnified view of ROI. (k) plots the quantitative profile along line $l_1$ and $l_2$ respectively. Scale bars in (h) and (j2) denote 14 $\upmu$m. \label{fig_discussion1}}
\end{figure} 
Since deviation of LED positions also introduces unexpected low-frequency component, our method [Eq. (\ref{Eq. 2})] is more robust to LED misalignment compared with conventional reconstruction algorithms. We performed verification on simulated data as shown in Fig. \ref{fig_discussion1}. The ideal distance between two adjacent LEDs is 4 mm, and we added random shift to each LED position with a maximum amplitude of 1 mm as demonstrated in Fig. \ref{fig_discussion1} (a). Such randomly shifted LEDs typically occur to customized LED arrays. Fig. \ref{fig_discussion1} (c,d) shows the reconstructed amplitude and Fourier spectrum using AS-EPRY and FPM-WSI. The reconstruction of FPM-WSI is quite similar to the ground truth in Fig. \ref{fig_discussion1} (b), providing a neat background distribution. In contrast, the quality of AS-EPRY reconstruction is severely degraded by wrinkle artifacts. We also calculated the values of SSIM and PSNR based on the ground truth to quantitatively evaluate the reconstruction performance. As listed in Fig. \ref{fig_discussion1} (e), FPM-WSI obtains a higher score than AS-EPRY in terms of both criteria. Results of massive simulations on 500 images with different degree of LED positional shifting are plotted in Fig. \ref{fig_discussion1} (f1) for PSNR and Fig. \ref{fig_discussion1} (f2) for SSIM, which equally suggest that FPM-WSI suffers less from the deviation of LED positions. We compared FPM-WSI with other three state-of-the-art methods, adaptive step-size FPM \cite{zuo2016adaptive}, ADMM-FPM \cite{wang2022fourier}, and momentum-PIE \cite{maiden2017further}, which can be referred to \textbf{Supplementary 6}. Additional simulation studies are also included regarding noise interference and LED intensity fluctuations, and the superiority of FPM-WSI has also been verified.

Given the reduced dependency of FPM-WSI on precise LED positioning, it becomes feasible to implement FPM with a squared LED array and even number of LEDs. As depicted in Fig. \ref{fig_discussion1} (g), we collected the LR images of USAF target only using 12×12 LEDs of the array. Obvious vignetting effect can be found with many half-bright and half-dark images. In this case, placed right above the slide is not the central LED but the middle area between two adjacent LEDs making the aligning of LED array more difficult. Fig. \ref{fig_discussion1} (j1,j2) shows the reconstructed amplitude of AS-EPRY and FPM-WSI with their corresponding magnified view of ROI. Despite the great challenge associated with LED alignment, FPM-WSI obtains a full-FOV reconstructed image with enhanced resolution compared with the raw data in Fig. \ref{fig_discussion1} (h).
According to the quantitative plots illustrated in Fig. \ref{fig_discussion1} (k), the feature of group 9, element 3 on the target can be clearly resolved. The result of AS-EPRY, however, is significantly distorted due to the vignetting effect as well as potential misalignment of LED positions.

\subsection{Pupil function recovery}

\begin{figure}
\centering
\includegraphics[width = 0.9\textwidth,trim = 0 330 0 0,clip]{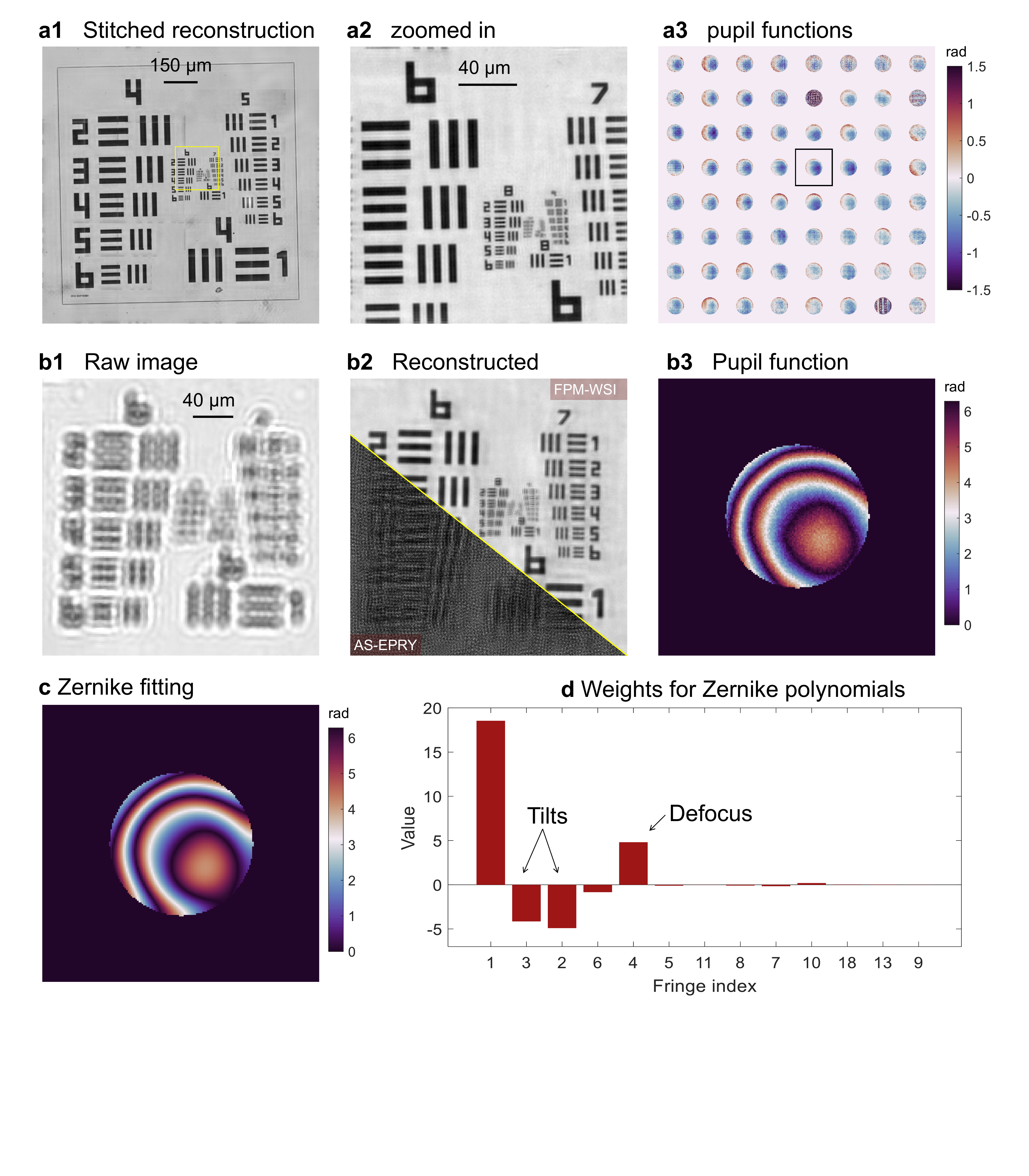}
\caption{Embedded pupil function recovery and digital refocusing for FPM reconstruction. (a1) Stitched reconstruction for a USAF target consisting of 16 image segments. (a2) Zoomed-in image of the region marked by the yellow box. (a3) Reconstructed spatially varying pupil functions for each segment. (b1) Central brightfield raw image of a defocused USAF target. (b2) Reconstructed results using AS-EPRY and FPM-WSI. (b3) and (c) are reconstructed pupil function corresponding to (b1) and its Zernike smoothed output. (d) plots the first 13 coefficients of the Zernike polynomial listed by fringe index. \label{fig_refocus}}
\end{figure}

\textbf{Local aberration recovery:} The use of objective lenses in FPM inherently introduces aberrations. Conventional reconstruction methods address this by incorporating updates to the pupil function within iterative phase retrieval processes, enabling correction of these aberrations. The fidelity function of FPM-WSI gives mathematical reciprocity to both $\mathbf{P}$ and $\mathbf{U}$, and thus their positions can be exchanged. We can then recover the pupil function by calculating the derivative w.r.t. $\mathbf{P}$ from Eq. (\ref{Eq. 2}) during the optimization. As shown in Fig. \ref{fig_refocus} (a1), we divided the full-FOV raw data of USAF target into 64 small segments, reconstructed them respectively and finally stitched reconstructions of all image segments. Each of them can be assigned a specific aberration-correction pupil function which is assumed to be unchanged in that region. The recovery of spatially varying pupil function for each image segment is demonstrated in Fig. \ref{fig_refocus} (a3). Fig. \ref{fig_refocus} (a2) shows the magnified view of one image segment, and its corresponding result of pupil recovery is marked by a yellow box in Fig. \ref{fig_refocus} (a3).

\textbf{Computational refocusing:} Sometimes, the high magnification objectives used in conventional WSI systems cannot capture precisely focused images within a narrow focal range due to the three-dimensional aspect or thick nature of slides. Layered scanning along the $z$-axis plane, also known as "$z$-stack", has become an increasingly prevalent practice to deal with this, by which a series of images are captured at multiple focal planes and then digitally combined to form a clearly focused composite. The number of scanning layers should be determined according to the evaluation of features in each image segment. This optimization of image capture through broadening of focus is time-consuming, and leads to the overall reduction of digitization speed considering the subsequent image stitching. 

Sample defocus is equivalent to introducing a defocus phase factor (4th-order Zernike function) to the pupil plane. Therefore, we can safely regard defocus as a special type of optical aberration (that is, a defocus aberration). The depth of focus (DOF) of the imaging system can be extended beyond that of the objective lens. In the initial FPM implementation \cite{zheng2013wide}, digital refocusing was dependent on adding a predefined defocused wavefront to correct the pupil function. When the defocus distance is unknown, ergodic reconstruction is necessary followed by identifying the sharpest image either manually or using softwares. For a tilted sample, this approach achieves acuity for different regions of the entire image, and stitches these focused regions together to complete refocusing. Thereafter, improved algorithms never broke the constraint that the defocus distance should be known. Even some network-based methods require training to learn a prior over $z$-slices \cite{bouchama2023fourier} or interpolation along the $z$-axis \cite{zhou2023fourier}. As shown in Fig. \ref{fig_refocus} (b1), the USAF target is placed off the focus plane with an unknown defocus distance. After the FPM-WSI reconstruction, both high-resolution amplitude in Fig. \ref{fig_refocus} (b2) and pupil function in Fig. \ref{fig_refocus} (b3) can be directly recovered. Zernike fitting image of the pupil function given in Fig. \ref{fig_refocus} (c) has three $2\pi$ jumpings implying very large aberration values. The plot of Zernike coefficients denotes that defocus aberration and tilting aberration exist in the imaging system. The reconstruction of AS-EPRY fails without a priori defocus distance for aberration compensation as the algorithm falls into local minima due to large aberrations. Severe vignetting effect also hinders the recovery of both sample wavefront and pupil function.

\subsection{Resolution of FPM-WSI platform} 
To determine the design parameters of the illumination source, we manufactured a programmable 25×25 LED array in advance and experimentally examined its limit resolution on the data of USAF target with green channel illumination. Considering the size of selected LED unit (3.5 mm × 3.5 mm) and the manufacturing technology, the distance between two adjacent LEDs was set as 4 mm. The experimental conditions basically remained unchanged as in section 2, except that a larger number of LEDs were used to provide an extended synthetic NA (the theoretical value is approximately 0.8, defined by the sum of objective NA and illumination NA). 

\begin{figure}
\centering
\includegraphics[width = 0.9\textwidth,trim = 0 720 0 0,clip]{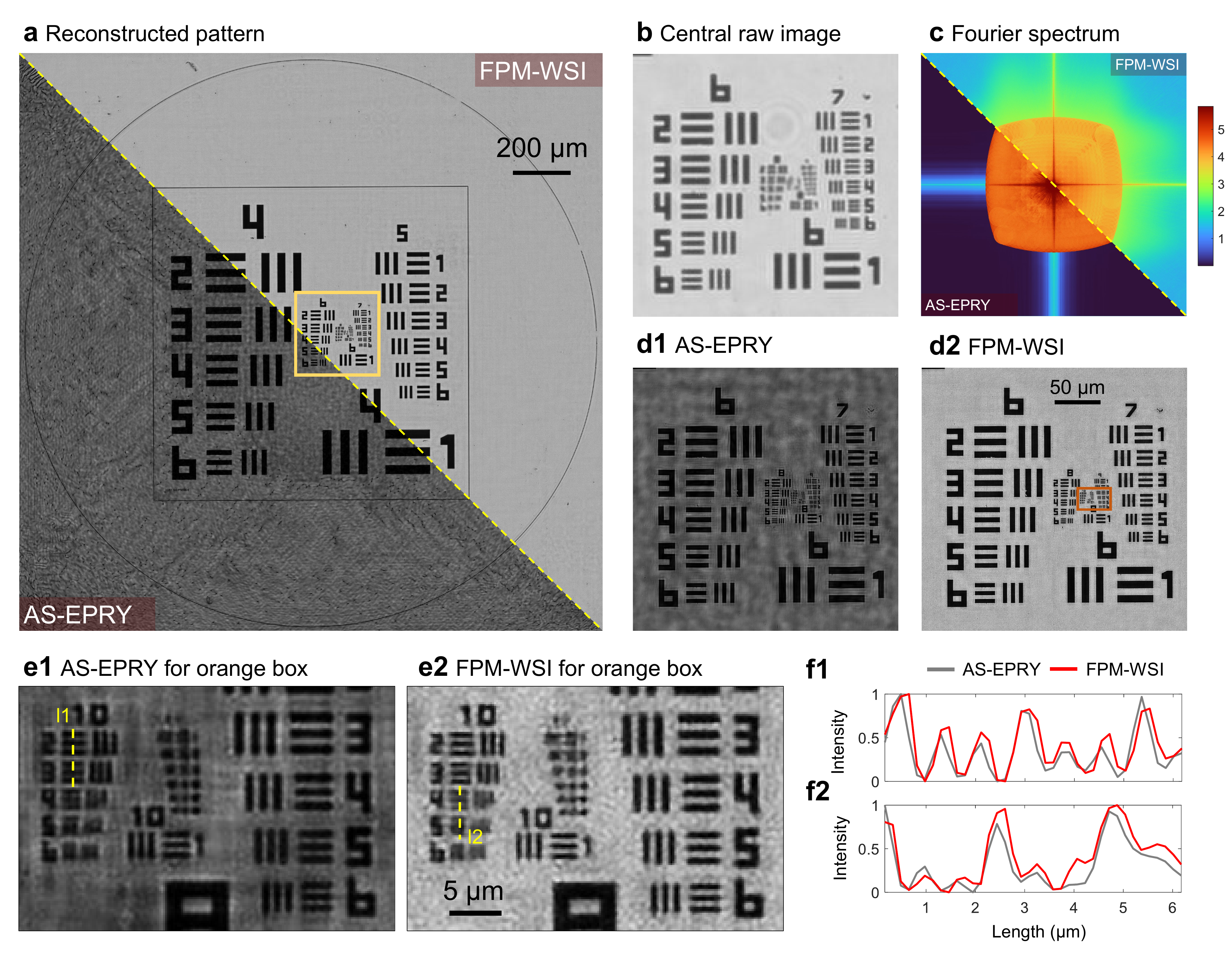}
\caption{Reconstruction of USAF resolution target with $25 \times 25$ LEDs. (a) Direct full-FOV reconstruction using AS-EPRY and FPM-WSI. (b) Magnified view of group 6-11 elements in the central brightfield raw image, marked by yellow box in (a). (c) Fourier spectrum for AS-EPRY and FPM-WSI reconstruction. (d1,d2) are reconstructed amplitudes of AS-EPRY and FPM-WSI corresponding to (b). (e1,e2) are magnified images of the region marked by the orange box in (d2). (e) plots the intensity profiles along the dashed lines in (e1,e2). (f1) and (f2) plot the intensity profiles along lines $l_1$ and $l_2$, respectively. \label{fig_resolution}}
\end{figure} 

According to Fig. \ref{fig_resolution} (b), the line structures of group 7, element 5 can be clearly identified with the illumination of central LED. Fig. \ref{fig_resolution} (d1,d2) compare the reconstructed amplitude of AS-EPRY and FPM-WSI for the marked region in Fig. \ref{fig_resolution} (a). Fig. \ref{fig_resolution} (e1,e2) demonstrate the corresponding magnified view of elements in group 10. After the synthetic aperture was completed, both of them obtained a great improvement of resolution with resolvable manifestation of group 10, element 3 (388 nm half-pitch resolution), as evidenced by Fig. \ref{fig_resolution} (f). The achievable resolution of an FPM system is jointly determined by the illumination wavelength and synthetic NA. In this implementation, however, the practical synthetic NA is only about 0.7, which roughly equals to the value that can be achieved by 19×19 LEDs (${NA_\text{syn}}=0.68$). It is indicated that the LEDs exceeding the illumination NA of 0.6 (${NA_\text{obj}}=0.1$) can no longer provide effective information of the sample for their corresponding raw images. This is the reason why a programmable 19×19 LED array was designed for our FPM-WSI platform with a fixed illumination height of 70 mm. Under this configuration, the highest half-pitch resolution of gray images (with blue channel illumination) achieved by this platform reaches 336 nm.

Although AS-EPRY reaches a comparable level of reconstructed resolution with FPM-WSI, the full-FOV result is severely degraded by wrinkle artifacts due to vignetting. The middle region of the image is free of artifacts, but the background is distributed with uneven patches, as demonstrated in Fig. \ref{fig_resolution} (d1). This experiment offered substantial evidence for high experimental robustness and data fidelity of FPM-WSI in direct full-FOV reconstruction.

\section{Conclusions}
In this paper, we have presented an efficient computational framework for stitching-free FPM reconstruction. This method formulates the loss function of optimization on the feature domain of images, and thus the inverse problem of FPM reconstruction can be solved under the framework of feature extraction. The feature-domain error, after the processing of an optimizer, is backdiffracted through the optical system for the update of complex amplitude of sample and CTF. Such a design allows us to completely bypass the challenging vignetting effect in typical FPM systems without modification of LED layout. High-quality reconstruction no longer depends greatly on the procedure of blocking reconstruction and then stitching. Besides, the method effectively deals with deviation of LED positions and LED intensity fluctuations, which reduces the requirement of precise calibration of systematic errors. For some certain experimental conditions, which are tough for conventional algorithms to conduct successful reconstructions, our method can also obtain impressive performances, such as digital refocusing without a priori defocus distance and reconstruction of the data with a lower overlapping rate of spectrum.

We further developed a FPM-WSI platform based on this framework, which firstly applied stitching-free FPM to the field of whole slide imaging. We list several typical characteristics of the platform as follows: (1) high-speed data acquisition (within 4 s for a single slide); (2) automatic and batch processing of 4 slides; (3) extension for multiple imaging modes and techniques; (4) user-friendly workflow with optional colorization schemes; (5) low cost of hardware without requirement of GPU-based devices. The reported platform is expected to promote the widely accepted application of FPM in the field of digital pathology. We believe that the improvement of hardware capabilities can adapt to more complex application scenarios. For example, the FOV of reconstruction demonstrated in Section 2 only occupies a small portion of the entire sample. The employment of cameras with larger sensor area will enable observation covering the whole range of slides, and provide essential and more comprehensive references for users' assessment. The time required for acquisition of each raw image can ascend to the limit value 2 ms using cameras with higher readout speed, which is particularly suited to intraoperative examination of pathological slides and facilitates prompt formulation of operation plans for clinicians. Furthermore, in applications of large-scale pathological imaging and biomedical research, designing a large-capacity storage and transportation device guarantees efficient operations.

More modifications and extensions on the framework can be implemented in the future. The existing framework utilizes the first-order gradient to extract edge features from images, addressing vignetting effect, noise signals and a series of systematic errors in the feature domain. While the detection of first-order edge is proved to be effective, the precious role that the second-order gradient of images plays in overcoming vignetting effect cannot be neglected. However, calculating the second-order gradient of images can inadvertently amplify noise signals. To isolate the valid features of images while suppressing the noise, we believe that training various band-limited edge filters through dictionary learning or the development of neural networks can be a good solution.These trained filters are designed to have a bandwidth that precisely matches the width of pupil function. Future works will include but are not limited to reflective FPM \cite{guo2015fourier,lee2019reflective}, near-field FPM \cite{zhang2019near} and tomographic FPM \cite{li2015separation,horstmeyer2016diffraction,zuo2020wide}. Imaging at a more macroscopic scale, such as remote sensing \cite{tian2023optical}, might also draw some meaningful inspirations from our method. 

\begin{backmatter}
\bmsection{Funding}
National Natural Science Foundation of China (NSFC) (No.12104500); Key Research and Development Projects of Shaanxi Province of China (No. 2023-YBSF-263) 

\bmsection{Acknowledgments}
An Pan thanks Jiurun Chen (Tsinghua University, China) for his contributions to the design of FPM-WSI platform, and assistant engineer Huiqin Gao (Xi’an Institute of Optics and Precision Mechanics, Chinese Academy of Sciences, China) for her constructive discussions on the redundant information model for FPM.

\bmsection{Disclosures}
The authors declare no competing financial interests.
\end{backmatter}


\bibliography{sample}






\end{document}